\documentclass[journal]{IEEEtran}
\usepackage{booktabs}
\usepackage{cite}
\usepackage{amsmath}
\usepackage{amsfonts}
\usepackage{algorithmic}
\usepackage{dblfloatfix}
\usepackage{graphicx}
\usepackage{textcomp}
\usepackage{caption}
\usepackage{subcaption}
\usepackage{atbegshi}
\usepackage{xstring}
\usepackage{flushend}
\usepackage[normalem]{ulem}
\usepackage{tikz}
\usepackage{multirow}
\usepackage{color}
\definecolor{forestgreen}{RGB}{0,139,69}
\usepackage{xcolor}
\definecolor{citecolor}{HTML}{0071bc}
\definecolor{SeaGreen4}{RGB}{0,205,102} 
\definecolor{SlateBlue}{RGB}{106,90,205} 
\definecolor{DarkRed}{RGB}{178,34,34} 
\usepackage[colorlinks, linkcolor=red,  anchorcolor=DarkRed, citecolor=SeaGreen4]{hyperref} 
\usepackage{makecell}
\usepackage{amssymb}
\usepackage{CJKutf8} 
\usepackage{mathrsfs}
\usepackage{soul}
\usepackage{colortbl}
\usepackage{pgfplotstable} 
\usepackage{pgfplots}
\pgfplotsset{compat=newest} 
\usepackage{pgffor} 
\usepackage{filecontents}
\usepackage{enumitem}
\usepackage{pifont} 

\newcommand{\x}{\mathbf{x}} 
\newcommand{\X}{\mathcal{X}} 

\newcommand{\nd}{{n}_{d}}  
\newcommand{\nv}{{n}_{v}} 
\newcommand{\nr}{{n}_{r}} 

\newcommand{\rn}{\mathbf{r}} 
\newcommand{\R}{\mathcal{R}} 
\newcommand{\re}{\mathbf{r}}  

\newcommand{\h}{H} 
\newcommand{\w}{W} 
\newcommand{\lgt}{L} 

\newcommand{\M}{\mathcal{M}} 

\newcommand{\fo}{\mathbf{F}} 
\newcommand{\Q}{\mathbf{Q}} 
\newcommand{\f}{\mathbf{F}_d} 
\newcommand{\fp}{\mathbf{F}'} 
\newcommand{\ff}{\mathbf{f}} 

\newcommand{\V}{\mathbf{V}} 

\newcommand{\Tp}{\mathbf{T}_{\text{prompt}}} 

\begin{document}

\title{Activating Associative Disease-Aware Vision Token Memory for LLM-Based X-ray Report Generation} 

\author{Xiao Wang, \emph{Member, IEEE}, Fuling Wang, Haowen Wang*, Bo Jiang*, \\ Chuanfu Li, Yaowei Wang, \emph{Member, IEEE}, Yonghong Tian, \emph{Fellow, IEEE}, Jin Tang
\thanks{$\bullet$ Xiao Wang, Fuling Wang, Haowen Wang, Bo Jiang, and Jin Tang are with Information Materials and Intelligent Sensing Laboratory of Anhui Province, Anhui University, Hefei 230601, China; Anhui Provincial Key Laboratory of Multimodal Cognitive Computation, Anhui University, Hefei 230601, China; School of Computer Science and Technology, Anhui University, Hefei 230601, China. (email: \{xiaowang, jiangbo, tangjin\}@ahu.edu.cn)} 
\thanks{$\bullet$ Chuanfu Li is with the First Affiliated Hospital of Anhui University of Chinese Medicine, Hefei 230022, China (email: licf@ahtcm.edu.cn)} 
\thanks{$\bullet$ Yaowei Wang is with Peng Cheng Laboratory, Shenzhen, China, and Harbin Institute of Technology, Shenzhen, China. (email: wangyw@pcl.ac.cn)}
\thanks{$\bullet$ Yonghong Tian is with Peng Cheng Laboratory, Shenzhen, China, and National Engineering Laboratory for Video Technology, School of Electronics Engineering and Computer Science, Peking University, Beijing, China. (email: yhtian@pku.edu.cn)} 
\thanks{* Corresponding author: Haowen Wang and Bo Jiang}
}

\markboth{ IEEE Transactions on ***, 2025 } 
{Shell \MakeLowercase{\textit{et al.}}: Bare Demo of IEEEtran.cls for IEEE Journals}

\maketitle

\begin{abstract}
X-ray image based medical report generation achieves significant progress in recent years with the help of the large language model, however, these models have not fully exploited the effective information in visual image regions, resulting in reports that are linguistically sound but insufficient in describing key diseases. In this paper, we propose a novel associative memory-enhanced X-ray report generation model that effectively mimics the process of professional doctors writing medical reports. It considers both the mining of global and local visual information and associates historical report information to better complete the writing of the current report. Specifically, given an X-ray image, we first utilize a classification model along with its activation maps to accomplish the mining of visual regions highly associated with diseases and the learning of disease query tokens. Then, we employ a visual Hopfield network to establish memory associations for disease-related tokens, and a report Hopfield network to retrieve report memory information. This process facilitates the generation of high-quality reports based on a large language model and achieves state-of-the-art performance on multiple benchmark datasets, including the IU X-ray, MIMIC-CXR, and Chexpert Plus. 
The source code of this work is released on \url{https://github.com/Event-AHU/Medical_Image_Analysis}. 
\end{abstract}


\IEEEpeerreviewmaketitle

\section{Introduction} 

\IEEEPARstart{X}{-ray} medical image based report generation targets describing the findings or impressions using natural language. This task can greatly alleviate the work pressure on doctors and reduce the waiting time for patients, providing a feasible method for empowering artificial intelligence in smart healthcare. Although the task has made considerable progress in recent years, there are still many issues, such as the difficulty in detecting key diseases and the challenge in modeling the association between contextual samples. The research and development of medical report generation models still have a long way to go.

\begin{figure}
\centering
\includegraphics[width=1\linewidth]{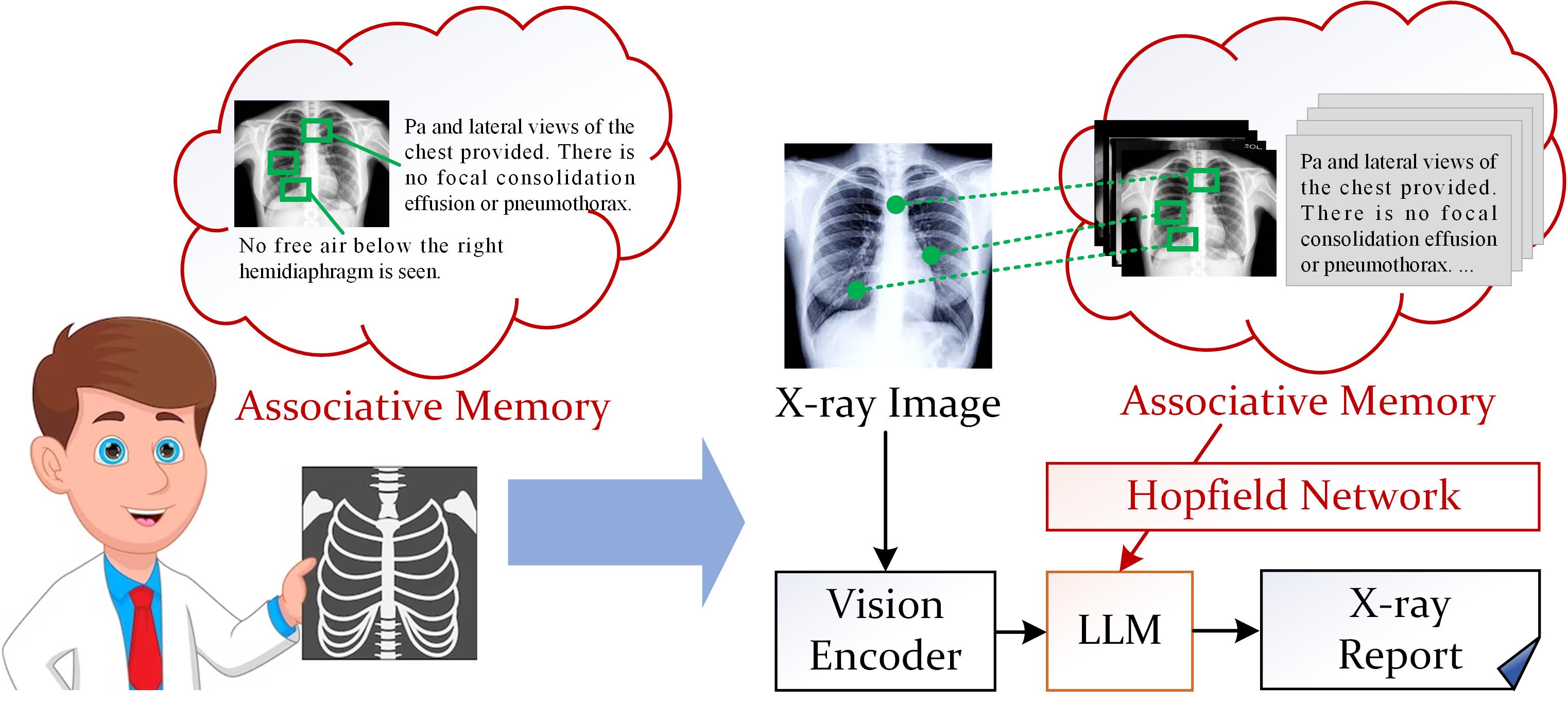}
\caption{A comparison between the process of professional doctors analyzing X-ray images and writing medical reports, and the proposed associative memory-enhanced X-ray LLM report generation framework.}
\label{fig:firstIMG}
\end{figure}

Recently, the Large Language Model (LLM) has drawn more and more attention in the community of artificial intelligence. It has demonstrated exceptional abilities in language understanding, reasoning, and generation. Some researchers already exploit the effectiveness of LLM in the X-ray based medical report generation, such as R2Gen-GPT~\cite{Wang2023R2GenGPT}, R2Gen-CSR~\cite{wang2024r2gencsr}, MambaXray-VL~\cite{wang2024CXPMRG}. Despite significant progress compared to traditional models, these models are still limited by the following issues: 
\textbf{Firstly}, current mainstream LLM-based report generation models focus on how to produce high-quality text at the linguistic level, but they struggle to accurately identify abnormal conditions, diseases, and other critical information in clinical diagnostic indicators. As a result, although the obtained medical reports may appear to be well-structured, they are actually difficult to address the practical problems. 
\textbf{Secondly}, existing medical report generation models, when considering contextual memory information, typically mine global image information, medical report data, etc., but rarely focus on fine-grained X-ray image information. However, these key visual details may be essential for producing accurate disease-related descriptions.
Thus, it is natural to raise the following questions: ``\textit{how can we utilize the local fine-grained visual cues and critical context information for X-ray medical report generation?}"

Considering that human medical experts have extensive experience in X-ray diagnosis, similarly, existing LLM-based models can understand and generate X-ray reports well, but medical experts also rely on recalling past cases to assist in diagnosing current situations. Therefore, we need to introduce an associative memory mechanism in the medical report generation process, to draw inspiration from a historical sample bank and generate more accurate medical report content. Fig.~\ref{fig:firstIMG} provides a comparison between human medical experts completing high-quality report writing through associative memory and the associative memory-enhanced LLM X-ray medical report generation proposed in this paper.

Inspired by these insights, we propose a novel associative memory augmented X-ray medical report generation framework via a large language model, termed AM-MRG. As shown in Fig.~\ref{fig:framework}, our framework contains two stages, i.e., the disease-aware visual token mining and the associative memory augmented X-ray medical report generation. 
Specifically, in the first stage, we extract the vision features of a given X-ray image using the Swin Transformer network~\cite{liu2021SwinTransformer}. Then, a Q-Former is adopted to augment the obtained vision features further and learn the disease queries, and a classification head is used for disease recognition. We utilize the GradCAM~\cite{jacobgilpytorchcam} to get the activation maps that reflect the key regions for disease recognition. Therefore, we can obtain massive disease-aware vision tokens and disease query features.

In the second stage, we take the visual tokens of each input X-ray image as the query feature and the disease-aware vision tokens as the knowledge base. A vision Modern Hopfield Network (MHN)~\cite{ramsauer2020hopfield} is introduced to achieve associative memory augmented feature learning for disease-aware X-ray medical report generation. Meanwhile, we also retrieve the report memory from the medical reports using a report MHN. A generation prompt is adopted to guide the report prediction of the large language model. Extensive experiments on three widely used benchmark datasets fully validated the effectiveness of the proposed AM-MRG framework.

To sum up, the main contributions of this paper can be summarized as follows: 

$\bullet$ We propose a novel associative memory augmented X-ray medical report generation framework based on a large language model, termed AM-MRG. The proposed modules have greatly improved the effectiveness of the X-ray medical reporter on clinical diagnostic indicators.

$\bullet$ We exploit the disease-aware visual token mining and report memory retrieval for associative memory construction based on Modern Hopfield Networks.

$\bullet$ Extensive experiments on three widely used benchmark datasets fully validated the effectiveness of our proposed associative memory augmented X-ray MRG framework.

\textit{The rest of this paper is organized as follows}: 
In section~\ref{relatedWorks}, we introduce the related works about the X-ray medical report generation, large language model, modern Hopfield network, and context sample retrieval. Then, we describe the proposed AM-MRG framework in Section~\ref{sec::framework}, with a focus on disease-aware visual token mining, report memory retrieval, associative memory augmented LLM for MRG, and details in the training and testing phrase. Then, we conduct experiments in Section~\ref{sec::experiments}, and conclude this paper in Section~\ref{sec::conclusion}.

\section{Related Work} \label{relatedWorks}

In this section, we will review the related works on the X-ray Medical Report Generation, Large Language Models, Modern Hopfield Network, and Context Sample Retrieval. More related works can be found in the following surveys~\cite{wang2024SSMSurvey, zhao2024RAGSurvey}.

\subsection{X-ray Medical Report Generation} 

Medical Report Generation (MRG)~\cite{Chen2020R2Gen, Tanida2023RGRG, Wang2023METransformer, huang2023kiut, hirsch2025medrat} is an important application area combining Natural Language Processing (NLP) and Computer Vision (CV), aiming to automate the generation of medical reports, particularly in fields such as radiology.   In MRG, models typically need to process two primary sources of information: visual information from medical images and linguistic information from existing medical literature or reports. 
R2Gen~\cite{Chen2020R2Gen} introduces a memory-driven Transformer for radiology report generation, using relational memory and memory-driven conditional layer normalization to enhance performance.    
RGRG~\cite{Tanida2023RGRG} uses a region-guided approach to detect and describe prominent anatomical regions, enhancing report accuracy and explainability while enabling new clinical applications. 
METransformer~\cite{Wang2023METransformer} introduces multiple learnable ``expert" tokens in the Transformer encoder and decoder and develops a metric-based ``expert" voting strategy to generate the final report.  
KiUT~\cite{huang2023kiut} model improves word prediction accuracy by learning multi-level visual representations through a U-Transformer architecture and adaptively extracting information using context and clinical knowledge.  
MedRAT~\cite{hirsch2025medrat} addresses the lack of paired image-report data during training by bringing relevant images and reports closer in the embedding space through auxiliary tasks such as contrastive learning and classification, enabling unpaired learning.  
HERGen~\cite{wang2024HERGen} effectively integrates longitudinal data from patient visits using group causal Transformers and enhances report quality with auxiliary contrastive objectives.  
PhraseAug~\cite{mei2024PhraseAug} introduces a phrase book as an intermediate modality, discretizing medical reports into key phrases, thereby facilitating fine-grained alignment between images and text.
CoFE~\cite{li2025CoFE} uses counterfactual explanations to learn truthful visual representations and fine-tunes pre-trained large language models (LLMs) with learnable prompts to generate semantically consistent and factually complete reports.    
However, existing studies are somewhat insufficient in handling complex disease information and ensuring the medical and logical coherence of generated reports.

\subsection{Large Language Models} 

In the field of MRG using Large Language Models (LLMs), recent research has focused on improving the integration of visual information and enhancing the accuracy of generated radiology reports.    
R2GenGPT~\cite{Wang2023R2GenGPT} employs a lightweight visual alignment module to align image features with the LLM's word embedding space, effectively processing image data and generating radiology reports.    
To address the limitations of existing methods in cross-modal feature alignment and text generation, Liu et al.~\cite{liu2024bootstrapping} leverages the powerful learning capabilities of LLMs, optimizing radiology report generation through domain-specific instance induction and a coarse-to-fine decoding process.  
HC-LLM~\cite{liu2024HCLLM} enhances the adaptability and performance of LLMs in radiology report generation tasks by utilizing historical diagnostic information and extracting both temporal shared and specific features from chest X-rays and diagnostic reports.    It applies three consistency constraints to ensure the accuracy of the reports generated and their consistency with disease progression.  
LLM-RG4~\cite{wang2024LLMRG4} leverages the flexible instruction-following capability and extensive general knowledge of LLMs.    By incorporating an adaptive token fusion module and a token-level loss weighting strategy, LLM-RG4 effectively handles diverse input scenarios and improves diagnostic precision. 
MambaXray-VL~\cite{wang2024CXPMRG} is a pre-trained large vision-language model proposed by Wang et al. which adopts Mamba for X-ray image encoding and Llama2 for report generation. 
Despite these advancements, existing methods have not combined the fine-grained visual modules with the powerful capabilities of LLMs.    Therefore, our AM-MRG takes a novel approach by innovatively combining these two aspects to achieve better performance in medical report generation.

\subsection{Modern Hopfield Network} 
The Modern Hopfield Network (MHN)~\cite{ramsauer2020hopfield} enhances the capability to store and retrieve patterns, similar to the attention mechanism in Transformers. This provides a new approach for integrating memory and attention mechanisms into deep learning models for medical report generation. By embedding Hopfield layers into the architecture, these networks can efficiently store and access visual features and intermediate representations, improving the ability to generate accurate and detailed radiology reports. This integration surpasses traditional fully connected, convolutional, and recurrent networks, offering advanced pooling, memory, association, and attention mechanisms.
To address the limitations of MHNs in memory storage and retrieval, Santos et al.~\cite{Santos2024SSHN} introduced the Hopfield-Fenchel-Young energy function, demonstrating that structured Hopfield Networks perform well in tasks such as multi-instance learning and text interpretation. Nicolini et al.~\cite{nicolini2024AssetAllocation} utilized MHNs to provide an efficient, scalable, and robust solution for asset allocation, showing that MHNs perform comparably or even better than deep learning methods such as LSTMs and Transformers, with faster training speeds and better stability.
Our AM-MRG is the first work to combine MHNs with medical report generation (MRG). By enhancing visual and textual representations through associative memory modules composed of MHNs, our approach has achieved impressive results.

\subsection{Context Sample Retrieving}  
Context Sample Retrieving is a technique for processing and generating natural language, particularly suited for large language models (LLMs) and generation tasks such as medical report generation. This technology aims to retrieve relevant context samples from a vast amount of data, allowing the model to reference these samples when generating medical reports, thereby improving the relevance and accuracy of the generated content. Chen et al.~\cite{chen2024learning} proposed a novel framework that treats the retrieval problem as a Markov decision process. Through policy optimization, the retriever can make iterative decisions to find the best example combinations for specific tasks. In-Context RALM~\cite{ram2023In-ContextRALM} significantly improves the performance of language models in generating medical reports by adding retrieved relevant documents to the input text without modifying the language model architecture. To address the challenges of few-shot hierarchical text classification, Chen et al.~\cite{chen2024HTC} built a retrieval database to identify examples related to the input text and employed an iterative strategy to manage multi-layer hierarchical labels. This method has shown excellent performance in few-shot HTC tasks. R2GenCSR~\cite{wang2024r2gencsr} retrieves positively and negatively related samples from the training set during the training phase and uses them to enhance feature representation and discriminative learning, thereby better guiding the LLM in generating accurate medical reports.

In our AM-MRG, the Context Sample Retrieving technique effectively integrates relevant medical visual and textual context information, providing the model with richer and more accurate references. This significantly improves the relevance and accuracy of the generated medical reports. The application of this technology notably enhances AM-MRG's performance in generating complex medical reports, ensuring the accuracy and clinical relevance of the report content.

\begin{figure*}
\centering
\includegraphics[width=0.9\linewidth]{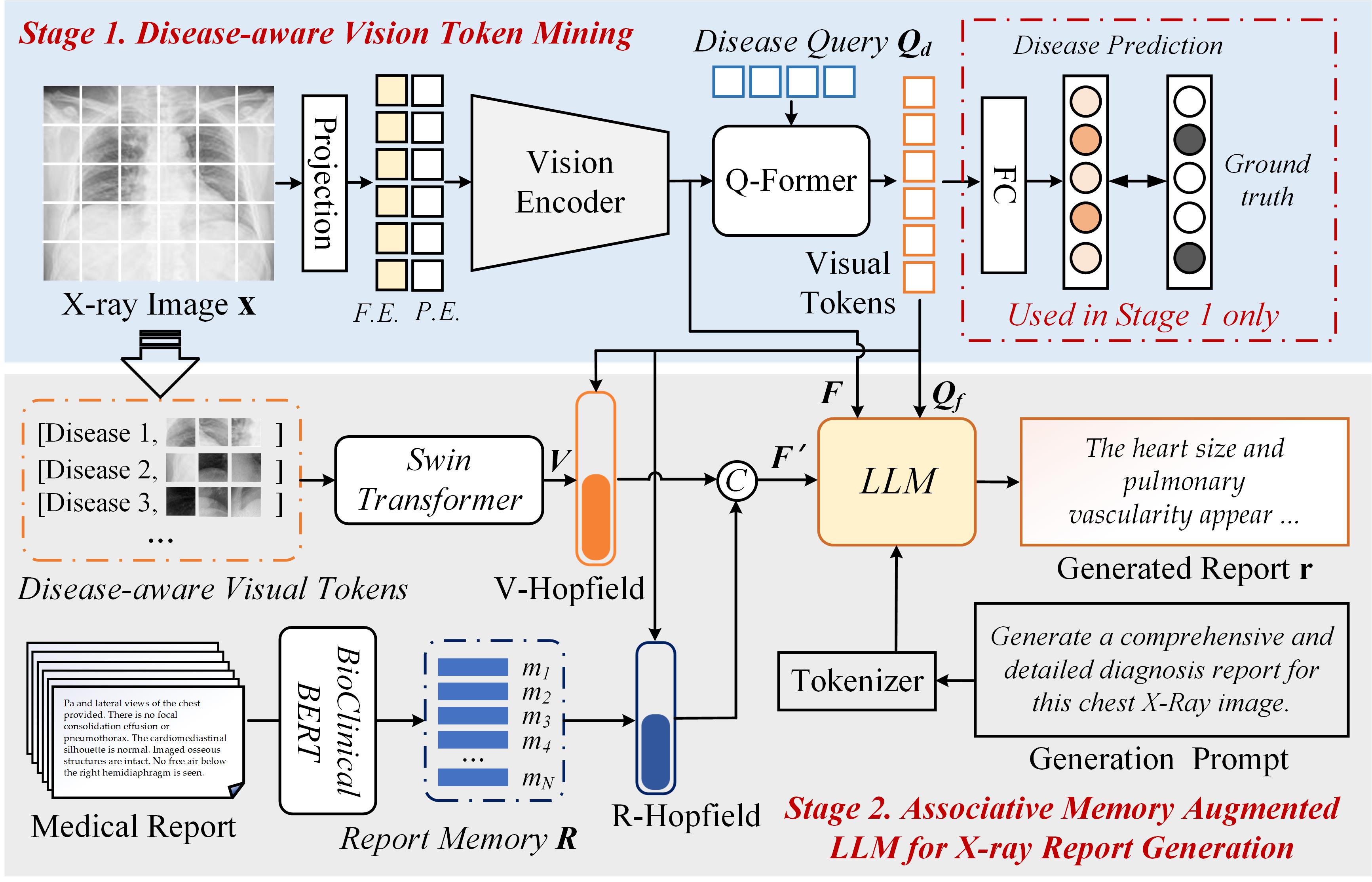}
\caption{An overview of our proposed Associative Memory augmented LLM for X-ray Medical Report Generation, termed AM-MRG. The first stage mainly mining the disease-aware visual tokens based on activation maps. The second stage attempts to augment the large language model-based X-ray medical reporter using associative memory. P.E. and F.E. are short for Position Encoding and Feature Embedding, respectively.
} 
\label{fig:framework}
\end{figure*}

\section{Methodology}  
\label{sec::framework}

\subsection{Overview} 
\label{sec:overview}

In this section, we introduce the proposed associative memory-augmented X-ray report generation framework~(AM-MRG). 
The framework generates an accurate and clinically relevant report $\rn \in \mathbb{R}^{\lgt}$ for a new X-ray image $\x \in \mathbb{R}^{\h \times \w \times 3}$ 
by leveraging a collection of existing X-ray images $\X = \{\x_{1}, \x_{2}, \dots, \x_{N}\}$ 
and their corresponding paired medical reports $\R = \{\rn_{1}, \rn_{2}, \dots, \rn_{N}\}$. 
The framework retrieves and refines relevant visual and textual information from this collection, dynamically enriching disease-specific representations that capture both semantic and contextual features. 
These representations are subsequently processed by a Large Language Model (LLM) $\M$ to produce the final diagnostic report $\rn$.

Generally speaking, our proposed AM-MRG framework consists of three main modules. 
In the first stage (Section~\ref{sec:token_mining}), 
disease query $\Q_d \in \mathbb{R}^{\nd \times 768}$ is introduced to extract disease-sepcific features $\f \in \mathbb{R}^{\nd \times 768}$ from each case image. 
Additionally, a classification task is employed to identify regions of interest (ROIs) that correspond to image patches associated with specific diseases. 
In the second stage (Section~\ref{sec:associative_memory}), 
two memory banks are constructed: 
a disease-aware visual bank $\V \in \mathbb{R}^{\nv \times 768}$, encoding region-specific visual features, 
and a report memory $\mathbf{R} \in \mathbb{R}^{\nr \times 768}$, encoding sentence-level semantic features from reports. 
These memory banks are leveraged via Vision Hopfield Network (V-Hopfield) and Report Hopfield Network (R-Hopfield) to refine the disease-specific features into enhanced representations $\fp \in \mathbb{R}^{(\nv + \nr) \times {4096}}$. 
Finally, in the report generation stage (Section~\ref{sec:report_gen}), 
the original features $\fo$ from the Vision Encoder, the query features $\Q_f$ from the Q-former, the enhanced features $\fp$, and the generation prompt together guide the LLM to produce the final report $\rn$.

\subsection{Disease-aware Vision Token Mining}
\label{sec:token_mining}
Given the input X-ray image $\x \in \mathbb{R}^{224 \times 224 \times 3}$, we first adopt the Swin-Transformer~\cite{liu2021SwinTransformer} for the feature extraction. Specifically, we divide the X-ray image into $196$ non-overlapping patches. Each of these patches is of a uniform size, measuring $16 \times 16$ pixels. Then, we project them into token representations using a projection layer and feed them into Transformer blocks for vision feature learning. The output $\textbf{F}$ will be fed into the Q-former network~\cite{li2023blip2} which further transforms them into feature representation $\Q_f \in \mathbb{R}^{14 \times 768}$. The used Q-former contains self-attention, cross-attention, and feed-forward layers, inspired by the BLIP2 model~\cite{li2023blip2}. 


Note that, we also input disease query $\Q_d \in \mathbb{R}^{\nd \times 768}$ into the Q-former to learn the embedding representation of each disease, where $\nd$ represents the number of disease categories, and $768$ denotes the query dimension. The cross-attention mechanism enables each disease query to interact with the encoded image features $\fo$, focusing on regions most relevant to specific diseases. A classification head is introduced to transform the visual features into 14 category labels defined in the MIMIC X-Ray dataset. The labels are obtained from the annotated medical reports. From stage 1 as shown in Fig.~\ref{fig:framework}, we can get the disease-aware query tokens and activation maps of input X-ray images which will be introduced in the following paragraphs.

To focus on disease-relevant areas in the X-ray image, we leverage GradCAM~\cite{jacobgilpytorchcam} activation maps \(\mathbf{M} \in \mathbb{R}^{H \times W}\) to identify Regions of Interest (RoI) that contribute most significantly to the classification task. For each patch \(\mathbf{r}_k\), the mean activation value is computed. A patch is retained as part of the RoI if its mean activation value exceeds a predefined threshold \(\tau\). Based on this selection, a masked version of the original image \(\mathbf{x}' \in \mathbb{R}^{H \times W \times 3}\) is generated, where non-RoI regions are masked to zero. The generation of \(\mathbf{x}'\) is defined as:
\begin{equation}
    \mathbf{x}'_{ij} = 
    \begin{cases} 
    \mathbf{x}_{ij} & \text{if } \frac{1}{|\mathbf{r}_{k}|} \sum_{(p, q) \in \mathbf{r}_{k}} \mathbf{M}_{pq} > \tau, \, (i, j) \in \mathbf{r}_{k}, \\
    0 & \text{otherwise},
    \end{cases}
\end{equation}
where 
\(\mathbf{r}_k\) denotes the \(k\)-th patch in the image, defined by its pixel indices \((p, q)\), 
and \(\mathbf{x}_{ij}\) represents the pixel value at location \((i, j)\) in the original image. 
Each patch contains \(|\mathbf{r}_k| = 16 \times 16\) pixels, and a patch is retained if the mean activation value of its corresponding region in the GradCAM activation map exceeds a predefined threshold \(\tau\).

This masking process ensures that only patches with significant activation values contribute to the final masked image \(\mathbf{x}'\), while other regions are suppressed. The resulting masked image \(\mathbf{x}'\) retains the spatial structure of the original image but focuses exclusively on the most critical disease-relevant regions. It is subsequently used in the encoding stage to construct the disease-aware visual bank, concentrating feature extraction on the preserved RoIs.

\subsection{Associative Memory Augmented Features}
\label{sec:associative_memory}

Two memory banks are constructed to encapsulate visual and textual knowledge. The \textit{disease-aware visual bank} $\V \in \mathbb{R}^{\nv \times 768}$ is built by encoding the masked X-ray images $\X' = \{\x'_{1}, \x'_{2}, \dots, \x'_{N}\}$, obtained from the previous stage, using a Swin Transformer~\cite{liu2021SwinTransformer}. This process produces region-specific feature tokens corresponding to the RoIs identified in each image. The \textit{Report Memory} $\mathbf{R} \in \mathbb{R}^{\nr \times 768}$ is constructed by encoding the medical reports $\R = \{\rn_{1}, \rn_{2}, \dots, \rn_{N}\}$ using a Bio\_ClinicalBERT~\cite{Emily2019Bio_ClinicalBERT} model. Each report is tokenized into sentences and processed into semantically rich sentence-level representations. These memory banks encapsulate comprehensive visual and textual knowledge, providing a robust foundation for dynamic retrieval and feature enhancement.

\begin{figure}
    \centering
    \includegraphics[width=0.8\linewidth]{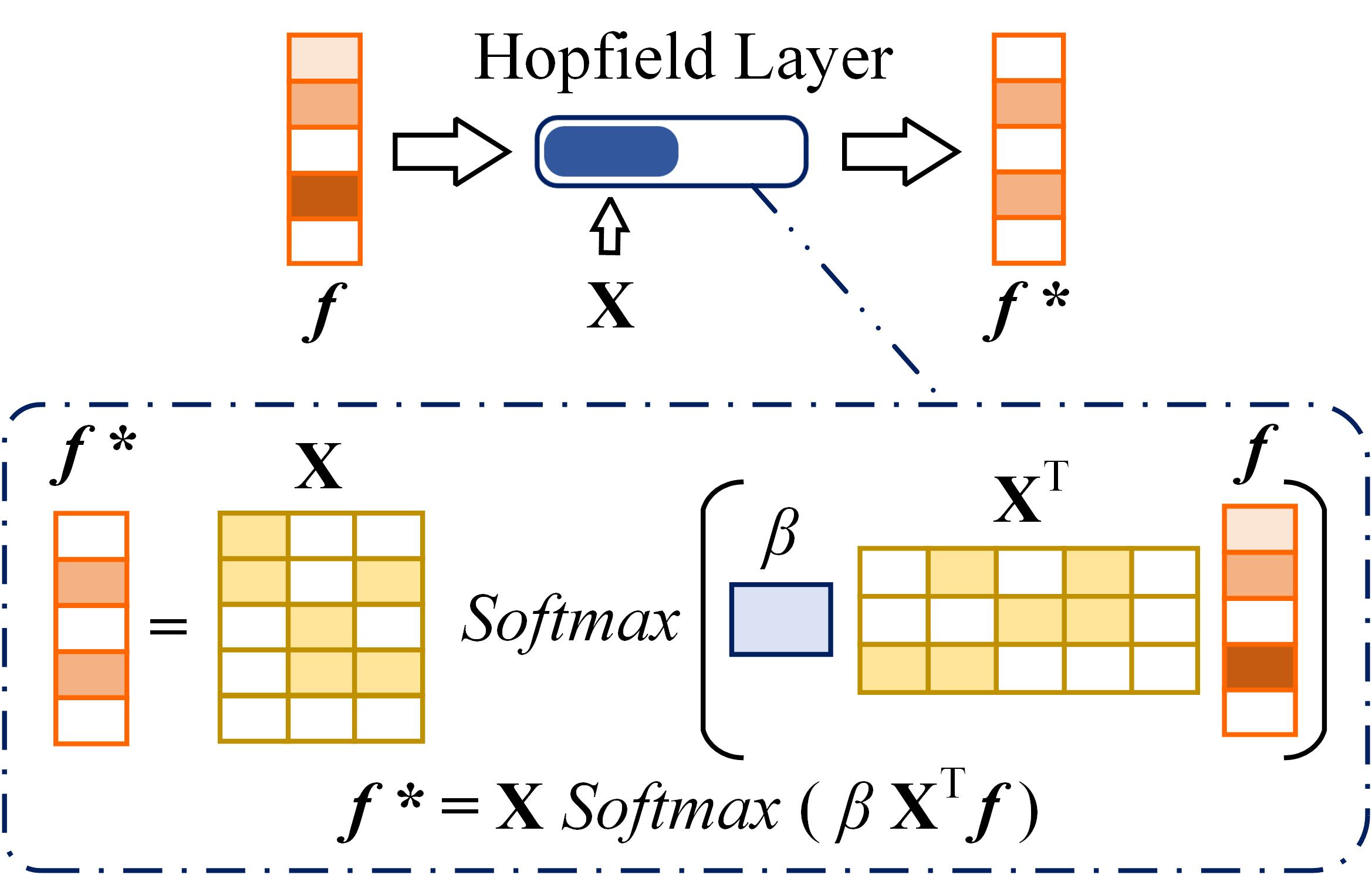}
    \caption{Illustration of compute procedure of Modern Hopfield Network (MHN).} 
    \label{fig:HopfieldNet}
\end{figure}

To augment the disease-specific features $\f \in \mathbb{R}^{14 \times 768}$ generated in the previous stage, we employ two modern Hopfield networks~\cite{ramsauer2020hopfield} for feature retrieval and enhancement, as shown in Fig.~\ref{fig:HopfieldNet}. 
Given a query feature $\ff_i \in \mathbb{R}^{768}$ and a memory bank $\mathbf{X} \in \mathbb{R}^{N \times 768}$ (where $\mathbf{X}$ can be either $\V$ or $\mathbf{R}$), 
the Hopfield network maps the query to an updated feature $\ff^*_i \in \mathbb{R}^{4096}$ by iteratively minimizing the following energy function:
\begin{equation}
    E(\ff_i^*) = \|\ff_i^* - \ff_i\|^2 - \beta \log \sum_{j=1}^N \exp \left( \frac{\ff_i^{*\top} \mathbf{m}_j}{\sqrt{d}} \right),
\end{equation}
where $\ff^*_i$ is initialized to $\ff_i$, $\mathbf{m}_j \in \mathbb{R}^{4096}$ represents the $j$-th stored feature in the memory bank, 
$\beta$ controls the sharpness of the similarity distribution, and $d$ normalizes the feature dimensions.

The updated feature $\ff^*_i$ is computed iteratively using gradient-based optimization:
\begin{equation}
    \ff_i^{\mathop{*}^{t+1}} = \ff_i^{\mathop{*}^t} - \eta \frac{\partial E(\ff_i^{*})}{\partial \ff_i^{*}},
\end{equation}
where the gradient is given by:
\begin{equation}
    \frac{\partial E(\ff^*_i)}{\partial \ff^*_i} = 2 (\ff^*_i - \ff_i) - \beta \sum_{j=1}^N \alpha_j \mathbf{m}_j,
\end{equation}
and:
\begin{equation}
    \alpha_j = \frac{\exp \left( \frac{\ff_i^{*\top} \mathbf{m}_j}{\sqrt{d}} \right)}{\sum_{k=1}^N \exp \left( \frac{\ff_i^{*\top} \mathbf{m}_k}{\sqrt{d}} \right)}.
\end{equation}

After convergence, the Hopfield network outputs the updated feature $\ff^*_i$, which represents the enhanced representation of the query based on the memory bank. 
We denote this operation as:
\begin{equation}
    \ff^*_i = \text{Hopfield}(\ff_i, \mathbf{X}),
\end{equation}
where $\mathbf{X}$ is the memory bank used for retrieval.

V-Hopfield for the \textit{Disease-aware Visual Bank} $\V$ and R-Hopfield for the \textit{Report Memory} $\mathbf{R}$ align input features with stored knowledge through energy minimization. 
The refined visual and textual features are computed as:
\begin{equation}
    \ff^*_{v, i} = \text{V-Hopfield}(\ff_i, \V; \theta_v), \quad 
    \ff^*_{r, i} = \text{R-Hopfield}(\ff_i, \mathbf{R}; \theta_r).
\end{equation}
These outputs are concatenated to form the final enhanced representation:
\begin{equation}
    \fp_i = \text{Concat}(\ff^*_{v, i}, \ff^*_{r, i}) \in \mathbb{R}^{2 \times 4096}.
\end{equation}
The complete output for all disease queries is:
\begin{equation}
    \fp = \{\fp_1, \fp_2, \dots, \fp_{14}\} \in \mathbb{R}^{(\nd \times 2) \times 4096}.
\end{equation}
This enhanced representation integrates both visual and semantic contexts, enabling robust downstream report generation.

\subsection{LLM-driven Report Generation}
\label{sec:report_gen} 
In this stage, the original features $\fo$ from the vision encoder, the query features $\Q$ from the Q-former, the enhanced feature representation $\fp$, and a predefined generation prompt together guide the LLM $\M$ in producing the final medical report.

To ensure the generation of accurate and contextually relevant medical reports, a specific generation prompt is designed:
\textit{Generate a comprehensive and detailed diagnosis report for this chest X-ray image.}
This prompt provides explicit instructions to the LLM, defining the objective of the generation task. 
The prompt is tokenized using the LLM’s tokenizer, resulting in a sequence of token embeddings $\Tp$. 
Meanwhile, the enhanced feature representation $\fp$ is processed to align with the LLM's input format, ensuring that the disease-relevant semantic features and the textual context are effectively integrated.

The LLM $\M$ combines the original features $\fo$ from the Vision Encoder, the query features $\Q$ from the Q-former, the enhanced feature representation $\fp$ 
encapsulating disease-specific visual and textual knowledge, and the token embeddings of the generation prompt $\mathbf{T}_{\text{prompt}}$, providing the textual context for the task.
Through its multi-modal processing layers, the LLM dynamically integrates these inputs and generates the final report $\rn$ in natural language:
\begin{equation}
    \rn = \M(\fo, \Q, \fp; \mathbf{T}_{\text{prompt}}),
\end{equation}
where $\rn$ represents the diagnosis report for the input X-ray image. 
The integration of $\fo$, $\Q$, $\fp$ and $\Tp$ ensures that the generated report reflects the underlying disease features while maintaining coherence and contextual relevance.

\subsection{Loss Function} 
\label{sec:train}
The proposed framework is trained in two distinct stages: 
Stage 1 optimizes the disease query features for extracting disease-specific representations, and Stage 2 fine-tunes the Modern Hopfield Networks for feature enhancement and retrieval. During the testing phase, the trained disease query and Hopfield Networks are fixed to generate enhanced features for report generation. 

In \textbf{Stage 1}, the focus is on optimizing the disease query $\Q \in \mathbb{R}^{14 \times 768}$ to extract disease-specific features $\f \in \mathbb{R}^{14 \times 768}$ from X-ray images. This is achieved through a multi-label classification task, which ensures that each disease query captures features relevant to a specific disease, and an orthogonality loss, which enforces distinctiveness among queries. The classification loss is defined as:
\begin{equation}
    \mathcal{L}_{\text{class}} = - \sum_{j=1}^{14} \left( y_j \log \hat{y}_j + (1 - y_j) \log (1 - \hat{y}_j) \right),
\end{equation}
where $y_j$ is the ground-truth label for the $j$-th disease and $\hat{y}_j$ is the predicted probability. 
The trained disease query features are fixed after Stage 1 to generate the initial disease-specific representations $\f$.

In \textbf{Stage 2}, the goal is to optimize the Hopfield Networks (V-Hopfield and R-Hopfield) for feature retrieval and enhancement using memory banks $\V$ and $\R$. The input to this stage is the disease-specific representation $\f$ from Stage 1. The enhanced feature representation $\fp$ is generated by the Hopfield Networks and used to guide the frozen LLM for report generation. The training in this stage employs an autoregressive loss inspired by R2GenGPT~\cite{Wang2023R2GenGPT}:
\begin{equation}
    \mathcal{L}_{\text{gen}} = - \sum_{t=1}^{L} \log P(\re_t \mid \re_{<t}, \fo, \Q, \fp; \Tp),
\end{equation}
where $\re_t$ is the $t$-th token of the ground-truth report $\re$, $\fo$ represents the original features from the Vision Encoder, $\Q$ denotes the query features from the Q-former, $\fp$ is the enhanced feature representation, and $\Tp$ represents the tokenized generation prompt.

During the \textbf{testing phase}, the trained disease query $\Q$ is fixed. The process begins with extracting disease-specific representations $\f$ from the input X-ray image $\x$ using $\Q$. The fixed Hopfield Networks use $\f$ to retrieve and enhance features from the memory banks $\V$ and $\R$, generating $\fp$. Finally, the original features $\fo$ from the vision encoder, the query features $\Q$ from the Q-former, the enhanced features $\fp$, and the generation prompt $\Tp$ are input to the LLM $\M$, which generates the diagnosis report $\rn$.






\section{Experiments} \label{sec::experiments}

\subsection{Dataset and Evaluation Metric}  

In our experiments, three benchmark datasets are selected for the evaluation, including \textbf{IU X-ray}~\cite{demner2016iuxray}, \textbf{MIMIC-CXR}~\cite{johnson2019mimicCXR}, \textbf{Chexpert Plus}~\cite{chambon2024CheXpertPLUS} dataset. A brief introduction to these datasets are given below. 

\noindent $\bullet$ \textbf{IU X-Ray}~\cite{demner2016iuxray} is a collection of chest X-ray images along with their corresponding diagnostic reports. The dataset comprises 7,470 pairs of images and reports. Each report consists of the following sections: impression, findings, labels, comparison, and indication. To ensure a equitable evaluation, we adhered to the same dataset partitioning scheme as used by R2GenGPT~\cite{Wang2023R2GenGPT}, allocating the dataset into training, testing, and validation subsets in a ratio of 7:1:2.

\noindent $\bullet$ \textbf{MIMIC-CXR}~\cite{johnson2019mimicCXR} is an extensive, publicly available chest radiology report dataset sourced from the Beth Israel Deaconess Medical Center in Boston, USA.  Encompassing 227,835 radiological studies, the dataset comprises a total of 377,110 chest X-ray images.  Designed to bolster advancements in the fields of computer vision and machine learning, MIMIC-CXR is poised to facilitate the development of automated medical image analysis technologies, aiding in the training of algorithms to detect and categorize pathological features within chest X-rays. To maintain a level playing field in our comparison, we adopted the same dataset partitioning approach as R2GenGPT~\cite{Wang2023R2GenGPT}, allocating 270,790 samples for model training, and assigning 2,130 and 3,858 samples to the validation and test sets, respectively.

\noindent $\bullet$ \textbf{Chexpert Plus}~\cite{chambon2024CheXpertPLUS} is a novel radiology dataset designed to improve the scale, performance, robustness, and equity of machine learning models in the field of radiology. It encompasses 223,228 chest radiographs, 187,711 corresponding radiology reports, de-identified demographic data for 64,725 patients, 14 chest pathology labels, and RadGraph~\cite{jain2021RadGraph} annotations. We adopted the dataset partitioning strategy proposed by ~\cite{wang2024CXPMRG} to ensure the fairness of our evaluation.

To evaluate our AM-MRG model, we employ widely recognized \textbf{NLG} (Natural Language Generation) metrics: \textbf{BLEU}~\cite{papineni_bleu_2002}, \textbf{ROUGE-L}~\cite{lin_rouge_2004}, and \textbf{METEOR}~\cite{banerjee_meteor_2005}, \textbf{CIDEr}~\cite{vedantam_cider_2015}. BLEU assesses the quality of text using n-gram matching; ROUGE-L examines text quality based on the longest common subsequence; and METEOR enhances BLEU by accounting for synonyms and word order; CIDEr assesses text using TF-IDF weighted n-gram matching, highlighting the significance of words. In addition, to assess the accuracy of clinical abnormality descriptions, we followed~\cite{Chen2020R2Gen, Tanida2023RGRG, nicolson2023improving}, use \textbf{CE} (Clinical Efficacy) metrics. Specifically, we extract labels from both the predicted and ground truth reports and compare the presence of key clinical observations to gauge the diagnostic accuracy of the generated reports. We evaluate the model's clinical efficacy using \textbf{Precision}, \textbf{Recall}, and \textbf{F1} scores. These CE metrics provide a comprehensive understanding of the model's performance in clinical applications.

\subsection{Implementation Details} 
The input X-ray images were resized to 224$\times$224 pixels and divided into 16$\times$16 patches, which were then fed into MambaXray-VL~\cite{wang2024CXPMRG} for visual feature extraction. To align with the visual feature dimensions output from Q-Former, the dimensions in both Hopfield~\cite{ramsauer2020hopfield} modules are set to 768. The feature dimension input to the LLM, after processing through various network layers, is 4096.  Unless otherwise specified, the LLM used for report generation is Llama2-7B~\cite{touvron2023llama2}, and the model used for report encoding is Bio\_ClinicalBERT~\cite{Emily2019Bio_ClinicalBERT}.  The key parameter \textit{Beta} in the Hopfield module is set to 4 .  The activation map mining method used is GradCAM~\cite{jacobgilpytorchcam}.  The number of visual disease regions extracted is capped at 500 per disease, with a total of 6943 regions for all 14 diseases combined.  The report memory size used is 6000, with the number of specific disease reports distributed according to their respective proportions. 
In stage 1, we set the learning rate as 5e-5 and adopt AdamW~\cite{loshchilov2017Adamw} optimizer for the training. In stage 2, we set the learning rate as 1e-4 and also adopt AdamW~\cite{loshchilov2017Adamw} optimizer for the training. 
In our research, we developed the model using PyTorch~\cite{paszke2019pytorch} and conducted training and testing on a server equipped with NVIDIA A800-SXM4-80GB GPU. More details can be found in our source code. 

\subsection{Comparison on Public Benchmark Datasets} 

\begin{table*}[]
\caption{ \textbf{NLG Metrics}. The symbol $\dagger$ indicates that we follow the R2Gen annotation using \textit{Findings} and evaluate with our method, as their report modifies the ground truth to an \textit{Impression} concatenated with \textit{Findings}. The best result is highlighted in bold, and the second-best result is underlined. }
\label{tab:NLG_Metrics_results}
\resizebox{\linewidth}{!}{
\begin{tabular}{c|l|c|ccccccc}
\hline \toprule [0.5 pt] 
\textbf{Dataset} & \textbf{Method} & \textbf{Publication} & \textbf{BLEU-1} & \textbf{BLEU-2} & \textbf{BLEU-3} & \textbf{BLEU-4} & \textbf{ROUGE-L} & \textbf{METEOR} & \textbf{CIDEr} \\ \hline

\multirow{10}{*}{\textbf{IU X-Ray}} 
 & R2Gen~\cite{Chen2020R2Gen} & EMNLP 2020 & 0.470 & 0.304 & 0.219 & 0.165 & 0.371 & 0.187 & - \\
 & R2GenCMN~\cite{Chen2021R2GenCMN} & ACL-IJCNLP 2021 & 0.475 & 0.309 & 0.222 & 0.170 & 0.375 & 0.191 & - \\
 & METransformer~\cite{Wang2023METransformer} & CVPR 2023 & \underline{0.483} & 0.322 & 0.228 & 0.172 & 0.380 & 0.192 & 0.435 \\
 & DCL~\cite{Li2023DCL} & CVPR 2023 & - & - & - & 0.163 & 0.383 & 0.193 & \underline{0.586} \\
 & R2GenGPT\textsuperscript{$\dagger$}~\cite{Wang2023R2GenGPT} & Meta Radiology 2023 & 0.465 & 0.299 & 0.214 & 0.161 & 0.376 & 0.219 & 0.542 \\
 & Token-Mixer~\cite{Yang2024Token-Mixer} & IEEE TMI 2024 & \underline{0.483} & \underline{0.338} & \underline{0.250} & \underline{0.190} & \textbf{0.402} & 0.208 & 0.482 \\
 & PromptMRG~\cite{Jin2024PromptMRG} & AAAI 2024 & 0.401 & - & - & 0.098 & 0.160 & \textbf{0.281} & - \\ 
 & Med-LMM~\cite{Liu2024Med_LMM} & ACM MM 2024 & - & - & - & 0.168 & 0.381 & 0.209 & 0.427 \\ 
 & SILC~\cite{Liu2024SILC} & IEEE TMI 2024 & 0.472 & 0.321 & 0.234 & 0.175 & 0.379 & 0.192 & 0.368 \\ 
 \cline{2-10} 
 & AM-MRG & Ours & \textbf{0.489} & \textbf{0.339} & \textbf{0.253} & \textbf{0.192} & \underline{0.384} & \underline{0.225} & \textbf{0.613} \\
 \hline \toprule [0.5 pt] 

\multirow{10}{*}{\textbf{MIMIC-CXR}} 
 & R2Gen~\cite{Chen2020R2Gen} & EMNLP 2020 & 0.353 & 0.218 & 0.145 & 0.103 & 0.277 & 0.142 & - \\
 & R2GenCMN~\cite{Chen2021R2GenCMN} & ACL-IJCNLP 2021 & 0.353 & 0.218 & 0.148 & 0.106 & 0.278 & 0.142 & - \\
 & METransformer~\cite{Wang2023METransformer} & CVPR 2023 & 0.386 & 0.250 & 0.169 & 0.124 & \textbf{0.291} & 0.152 & \textbf{0.362} \\
 & DCL~\cite{Li2023DCL} & CVPR 2023 & - & - & - & 0.109 & 0.284 & 0.150 & \underline{0.281} \\
 & R2GenGPT\textsuperscript{$\dagger$}~\cite{Wang2023R2GenGPT} & Meta Radiology 2023 & 0.408 & 0.256 & 0.174 & 0.125 & 0.285 & \underline{0.167} & 0.244 \\
 & Token-Mixer~\cite{Yang2024Token-Mixer} &IEEE TMI 2024 & \underline{0.409} & \underline{0.257} & \underline{0.175} & 0.124 & 0.288 & 0.158 & 0.163 \\
 & PromptMRG~\cite{Jin2024PromptMRG} & AAAI 2024 & 0.398 & - & - & 0.112 & 0.268 & 0.157 & - \\ 
 & Med-LMM~\cite{Liu2024Med_LMM} & ACM MM 2024 & - & - & - & \underline{0.128} & \underline{0.289} & 0.161 & 0.265 \\ 
 & AdaMatch-Cyclic~\cite{chen2024AdaMatch_Cyclic} & ACL 2024 & 0.379 & 0.235 & 0.154 & 0.106 & 0.286 & 0.163 & - \\
 \cline{2-10} 
 & AM-MRG & Ours & \textbf{0.426} & \textbf{0.271} & \textbf{0.187} & \textbf{0.136} & \textbf{0.291} & \textbf{0.174} & 0.261 \\ 
\hline  \toprule [0.5 pt] 

\multirow{10}{*}{\textbf{Chexpert Plus}} 
 & R2Gen~\cite{Chen2020R2Gen} & EMNLP 2020 & 0.301 & 0.179 & 0.118 & 0.081 & 0.246 & 0.113 & 0.077 \\
 & R2GenCMN~\cite{Chen2021R2GenCMN} & ACL-IJCNLP 2021 & 0.321 & 0.195 & 0.128 & 0.087 & 0.256 & 0.127 & 0.102 \\
 & XProNet~\cite{wang2022XProNet} & ECCV 2022 & 0.364 & 0.225 & 0.148 & 0.100 & 0.265 & \underline{0.146} & 0.121 \\
 & ORGan~\cite{hou2023ORGan} & ACL 2023 & 0.320 & 0.196 & 0.128 & 0.086 & 0.261 & 0.135 & 0.107 \\
 & R2GenGPT~\cite{Wang2023R2GenGPT} & Meta Radiology 2023 & 0.361 & 0.224 & 0.149 & \underline{0.101} & \underline{0.266} & 0.145 & \underline{0.123} \\
 & ASGMD~\cite{XUE2024ASGMD} &ESWA 2024 & 0.267 & 0.149 & 0.094 & 0.063 & 0.220 & 0.094 & 0.044 \\
 & Token-Mixer~\cite{Yang2024Token-Mixer} &IEEE TMI 2024 & \underline{0.378} & \underline{0.231} & \underline{0.153} & 0.091 & 0.262 & 0.135 & 0.098 \\
 & PromptMRG~\cite{Jin2024PromptMRG} & AAAI 2024 & 0.326 & 0.174 & - & 0.095 & 0.222 & 0.121 & 0.044 \\ 
 & R2GenCSR~\cite{wang2024r2gencsr} & arXiv 2024 & 0.364 & 0.225 & 0.148 & 0.100 & 0.265 & \underline{0.146} & 0.121 \\
 \cline{2-10} 
 & AM-MRG & Ours & \textbf{0.381} & \textbf{0.238} & \textbf{0.159} & \textbf{0.109} & \textbf{0.282} & \textbf{0.173} & \textbf{0.221} \\ 
\hline  \toprule [0.5 pt] 
\end{tabular}
}
\end{table*}

\begin{table}[]
\caption{ \textbf{CE Metrics} of AM-MRG in MIMIC-CXR. } 
\label{tab:CE_Metrics_Mimic_CXR_results}
\resizebox{\linewidth}{!}{
\begin{tabular}{l|c|ccc}
\hline \toprule [0.5 pt] 
\multirow{2}{*}{\textbf{Method}} & \multirow{2}{*}{\textbf{Publication}} & \multicolumn{3}{c}{\textbf{MIMIC-CXR}}  \\ \cline{3-5} 
 & & \textbf{Precison} &  \textbf{Recall} &  \textbf{F1}  \\ \hline
 R2Gen~\cite{Chen2020R2Gen} & EMNLP 2020  & 0.333 & 0.273 & 0.276   \\
 METransformer~\cite{Wang2023METransformer} & CVPR 2023  & 0.364 & 0.309 & 0.311 \\
 KiUT~\cite{huang2023kiut} & CVPR 2023  & 0.371 & 0.318 & 0.321 \\
 CoFE~\cite{li2025CoFE} & ECCV 2024 & 0.489 & 0.370 & 0.405  \\
 HERGen~\cite{wang2024HERGen} & ECCV 2024  & 0.415 & 0.301 & 0.317 \\
 SILC~\cite{Liu2024SILC} & IEEE TMI 2024  & 0.457 & 0.337 & 0.330 \\
 OaD~\cite{li2024OaD} & IEEE TMI 2024 & 0.364 & 0.382 & 0.372  \\
 
\hline 
 AM-MRG & Ours &  \textbf{0.555} & \textbf{0.429} & \textbf{0.484}   \\
\hline \toprule [0.5 pt] 
\end{tabular}
}
\end{table} 

\begin{table}[]
\caption{ \textbf{CE Metrics} of AM-MRG in Chexpert Plus. } 
\label{tab:CE_Metrics_CheXpert_Plus_results}
\resizebox{\linewidth}{!}{
\begin{tabular}{l|c|ccc}
\hline \toprule [0.5 pt] 
\multirow{2}{*}{\textbf{Method}} & \multirow{2}{*}{\textbf{Publication}} & \multicolumn{3}{c}{\textbf{Chexpert Plus}}  \\ \cline{3-5} 
 & & \textbf{Precison} &  \textbf{Recall} &  \textbf{F1}  \\ \hline
 R2Gen~\cite{Chen2020R2Gen} & EMNLP 2020  & 0.318 & 0.200 & 0.181   \\
 R2GenCMN~\cite{Chen2021R2GenCMN} & ACL 2021 & 0.329 & 0.241 & 0.231 \\
 XProNet~\cite{wang2022XProNet} & ECCV 2022  & 0.314 & 0.247 & 0.259 \\
 R2GenGPT~\cite{Wang2023R2GenGPT} & Meta-Rad. 2023  & 0.315 & 0.244 & 0.260 \\
 Zhu et al.~\cite{zhu2023utilizing} & MICCAI 2023 & 0.217 & 0.308 &  0.205 \\
 PromptMRG~\cite{Jin2024PromptMRG} & AAAI 2024  & 0.258 & 0.265 & 0.281 \\
 Token-Mixer~\cite{Yang2024Token-Mixer} & IEEE TMI 2024  & 0.309 & 0.270 & 0.288 \\
 
\hline 
 AM-MRG & Ours &  \textbf{0.396} & \textbf{0.318} & \textbf{0.336}   \\
\hline \toprule [0.5 pt] 
\end{tabular}
}
\end{table}


In our study, we rigorously compare our model with other state-of-the-art models across the three most commonly used datasets in the medical report generation field. This comparison includes classic cross-modal alignment models such as R2Gen~\cite{Chen2020R2Gen}, CMN~\cite{Chen2021R2GenCMN}, PromptMRG~\cite{Jin2024PromptMRG}, and XProNet~\cite{wang2022XProNet}. Additionally, we evaluate models like DCL~\cite{Li2023DCL} and ORGan~\cite{hou2023ORGan}, which enhance report generation by integrating additional medical knowledge graphs. Furthermore, we focus on models that leverage the powerful capabilities of Large Language Models (LLMs) to directly optimize the decoding process, including R2GenCSR~\cite{wang2024r2gencsr}, R2GenGPT~\cite{Wang2023R2GenGPT}, Med-LMM~\cite{Liu2024Med_LMM}, and AdaMatch-Cyclic~\cite{chen2024AdaMatch_Cyclic}. Lastly, we conduct an in-depth analysis of innovative models operating at the token and word levels, such as ASGMD~\cite{XUE2024ASGMD}, METransformer~\cite{Wang2023METransformer}, Token-Mixer~\cite{Yang2024Token-Mixer}, and SILC~\cite{Liu2024SILC}. The experimental results presented in the original papers of these models serve as a valuable reference for our comprehensive comparative analysis.

\noindent $\bullet$ \textbf{Results on NLG Metrics.~}
Table \ref{tab:NLG_Metrics_results} showcases the experimental outcomes of our method, AM-MRG, juxtaposed with traditional and contemporary approaches from the past two years, across three medical imaging report generation datasets. The table clearly demonstrates that AM-MRG outperforms its counterparts on the majority of NLG metrics, with a particularly stellar showing on the Chexpert Plus dataset, where it surpasses all other models across every metric.
On the IU X-Ray dataset, AM-MRG attains top scores with 0.489 for BLEU-1, 0.339 for BLEU-2, 0.253 for BLEU-3, 0.192 for BLEU-4, and 0.613 for CIDEr, each representing the highest among all competing methods. Notably, BLEU-4 sees a substantial enhancement of 19.2\% over R2GenGPT~\cite{Wang2023R2GenGPT}.
In the MIMIC-CXR dataset, AM-MRG secures the best results for all metrics except CIDEr, with scores of 0.426 for BLEU-1, 0.271 for BLEU-2, 0.187 for BLEU-3, 0.136 for BLEU-4, 0.291 for ROUGE-L, and 0.174 for METEOR. Additionally, BLEU-4 has improved by 8.8\% relative to R2GenGPT~\cite{Wang2023R2GenGPT}.
On the Chexpert Plus dataset, AM-MRG takes the lead in all metrics, posting values of 0.381 for BLEU-1, 0.238 for BLEU-2, 0.159 for BLEU-3, 0.109 for BLEU-4, 0.282 for ROUGE-L, 0.173 for METEOR, and 0.221 for CIDEr.
These results underscore the efficacy of our method in generating reports that closely align with reference reports, adeptly capturing crucial information while preserving linguistic fluidity. In comparison to other cutting-edge techniques, our method exhibits a notable edge, charting a new course for research in the domain of medical imaging report generation.

\noindent $\bullet$ \textbf{Results on CE Metrics.~} 
As depicted in Tables \ref{tab:CE_Metrics_Mimic_CXR_results} and \ref{tab:CE_Metrics_CheXpert_Plus_results}, we have evaluated the accuracy of our model, AM-MRG, in delineating clinical abnormalities by reporting the Clinical Efficacy (CE) metrics alongside seven other models across two distinct datasets. On the MIMIC-CXR dataset, AM-MRG yielded impressive results with Precision, Recall, and F1 scores of 0.555, 0.429, and 0.484, respectively. Notably, the Recall and F1 scores surpass those of the competing models, with the F1 score a remarkable 19.5\% higher than the second-best model, CoFE~\cite{li2025CoFE}. In the Chexpert Plus dataset, AM-MRG took the lead in all three metrics, achieving Precision, Recall, and F1 scores of 0.396, 0.318, and 0.336, respectively. The F1 score represented a significant 16.6\% improvement over the runner-up, Token-Mixer~\cite{Yang2024Token-Mixer}, signifying a substantial edge.

These CE metrics offer a nuanced perspective on the model's practical performance in clinical scenarios. The exemplary CE performance of AM-MRG on both datasets underscores its robustness and excellence. The significance of CE metrics is underscored by their dual emphasis on the linguistic fluency and coherence of the generated reports as well as their diagnostic precision, a critical factor in medical contexts. CE metrics guarantee that the produced medical imaging reports not only read coherently but also convey essential diagnostic insights for clinicians, thereby refining the accuracy and expediency of clinical decision-making processes.

\subsection{Ablation Study} 

\newcommand{\yes}{\textcolor{SeaGreen4}{\ding{51}}}
\newcommand{\no}{\textcolor{DarkRed}{\ding{55}}}

\begin{table*}[]
\caption{ Component analysis of the key modules in our framework on MIMIC-CXR and Chexpert Plus dataset.  \textbf{Visual}, \textbf{Report}, \textbf{B4}, \textbf{R-L}, \textbf{M}, \textbf{C}, \textbf{P}, \textbf{R} and \textbf{F1} represents V-Hopfield, R-Hopfield BLEU-4, ROUGE-L, METEOR, CIDEr, Precision, Recall and F1 respectively.} 
\label{tab:Ablation}
\resizebox{\linewidth}{!}{
\begin{tabular}{l|cc|ccccccc|ccccccc}
\hline \toprule [0.5 pt] 
\multirow{2}{*}{\textbf{Index}} & \multirow{2}{*}{\textbf{Visual}} & \multirow{2}{*}{\textbf{Report}} & \multicolumn{7}{c|}{\textbf{MIMIC-CXR}}   & \multicolumn{7}{c}{\textbf{Chexpert Plus}}  \\ \cline{4-17} 
 & & & \textbf{B4} & \textbf{R-L} &  \textbf{M}  & \textbf{C}  &  \textbf{P} &  \textbf{R} &  \textbf{F1} & \textbf{B4} & \textbf{R-L} &  \textbf{M}  & \textbf{C} &  \textbf{P} &  \textbf{R} &  \textbf{F1} \\ \hline
$\#01$ & \no & \no & 0.131 & 0.286 & 0.168 & 0.241 & 0.539 & 0.396 & 0.458 & 0.101 & 0.276 & 0.167 & 0.173 & 0.348 & 0.265 & 0.284\\
\hline 
$\#02$ & \yes & \no & 0.133 & \textbf{0.291} & 0.173 & \textbf{0.265} & 0.545 & 0.423 & 0.477 & 0.106 & 0.279 & 0.170 & 0.206 & 0.374 & 0.297 & 0.317 \\
$\#03$ & \no & \yes & 0.132 & 0.286 & 0.169 & 0.256 & 0.486 & \textbf{0.455} & 0.470 & 0.105 & 0.277 & 0.169 & 0.209 & 0.384 & 0.279 & 0.304 \\
\hline 
$\#04$ & \yes & \yes & \textbf{0.136} & \textbf{0.291} & \textbf{0.174} & 0.261 & \textbf{0.555} & 0.429 & \textbf{0.484} & \textbf{0.109} & \textbf{0.282} & \textbf{0.173} & \textbf{0.221} & \textbf{0.396} & \textbf{0.318} & \textbf{0.336} \\
\hline \toprule [0.5 pt] 
\end{tabular}
}
\end{table*}

\noindent $\bullet$ \textbf{Component Analysis.~}  
As shown in Table \ref{tab:Ablation}, we conducted experimental analyses on several key modules within the AM-MRG model in two datasets to determine their effectiveness.  The table clearly indicates that adding only the V-Hopfield module results in significant improvements in both NLG and CE metrics compared to the baseline method, which lacks both V-Hopfield and R-Hopfield modules. Similarly, adding only the R-Hopfield module also enhances performance in NLG and CE metrics over the Baseline.  When both modules are employed together, the improvements are even more pronounced.  In the MIMIC-CXR~\cite{johnson2019mimicCXR} dataset, the NLG metrics BLEU-4, ROUGE-L, METEOR, and CIDEr increased from 0.131, 0.286, 0.168, and 0.241 to 0.136, 0.291, 0.174, and 0.261, respectively, representing a substantial enhancement. In terms of CE metrics, the improvements were even more significant. In the Chexpert Plus dataset, Precision, Recall, and F1 scores rose from 0.348, 0.265, and 0.284 to 0.396, 0.318, and 0.336, respectively. Overall, the inclusion of these two modules resulted in significant performance boosts across both datasets, with particularly noteworthy improvements on the Chexpert Plus~\cite{chambon2024CheXpertPLUS} dataset in both NLG and CE metrics. These results demonstrate that the combination of V-Hopfield and R-Hopfield modules significantly enhances the performance of the AM-MRG model.

\begin{table}
\centering
\caption{Compare the effects of different numbers of vision CAM maps on outcomes.  }  
\label{tab:CAM_num}
\small 
\resizebox{\linewidth}{!}{ 
\begin{tabular}{c|c|cccc} 
\hline 
\toprule [0.5 pt] 
\textbf{Num}  &\textbf{Sum} &\textbf{BLEU-4}  &  \textbf{ROUGE-L} & \textbf{METEOR} & \textbf{CIDEr} \\
\hline 
 100 & 1400 & 0.129 & 0.282 & 0.160 & 0.209 \\
\hline 
 250 & 3500 & 0.130 & 0.284 & 0.167 & 0.239 \\
\hline 
 500 & 6943 & \textbf{0.136} & \textbf{0.291} & \textbf{0.174} & \textbf{0.261}\\
\hline 
 750 & 10193& 0.131 & 0.288 & 0.167 & 0.246 \\
\hline 
 1000 & 13263 & 0.129 & 0.283 & 0.159 & 0.212 \\
\hline 
\toprule [0.5 pt]  
\end{tabular}
} 
\end{table}

\noindent $\bullet$ \textbf{Analysis of varying numbers of disease-aware regions in V-Hopfield.~} 
As depicted in Table \ref{tab:CAM_num}, we have carried out a suite of experiments to examine the effect of varying the number of disease-aware regions on the outcomes. The table's \textbf{Num} column denotes the maximum count of disease-aware regions chosen for each disease type, acknowledging that less common diseases may have a limited number of regions. Conversely, the \textbf{Sum} column reflects the cumulative count of disease-aware regions across all diseases. 

In our experimental setup, the \textbf{Num} value was adjusted uniformly from 100 to 1000. The results revealed a steady improvement in performance as the \textbf{Num} value ascended from 100 to 500. The model's peak performance was observed at a \textbf{Num} value of 500, where the natural language generation (NLG) metrics BLEU-4, ROUGE-L, METEOR, and CIDEr achieved scores of 0.136, 0.291, 0.174, and 0.261, respectively. Beyond this point, as the \textbf{Num} value rose from 500 to 1000, there was a progressive decline in accuracy, suggesting that the optimal setting for \textbf{Num} is at 500 for peak model efficacy.

\begin{table}
\centering
\caption{Compare the effects of different numbers of medical reports on outcomes.  }  
\label{tab:Medical_report_num}
\small 
\begin{tabular}{c|cccc} 
\hline 
\toprule [0.5 pt] 
\textbf{Samples}  &\textbf{BLEU-4}  &  \textbf{ROUGE-L} & \textbf{METEOR} & \textbf{CIDEr} \\
\hline 
 500  & 0.132 & 0.285 & 0.164 & 0.233\\
\hline 
 1000 & 0.132 & \textbf{0.291} & 0.170 & 0.245\\
\hline 
 3000 & 0.130 & 0.287 & 0.166 & 0.222 \\
\hline 
 6000 & \textbf{0.136} & \textbf{0.291} & \textbf{0.174} & \textbf{0.261}\\
\hline 
 8000 & 0.133 & 0.288 & 0.165 & 0.256\\
\hline 
 10000 & 0.131 & 0.287 & 0.167 & 0.236\\
\hline 
\toprule [0.5 pt]  
\end{tabular}
\end{table}

\noindent $\bullet$ \textbf{Analysis of varying numbers of medical reports in R-Hopfield.~}
As shown in Table \ref{tab:Medical_report_num}, we conducted a series of experiments to investigate the impact of different quantities of medical reports on the final results. In the table, the column \textbf{Samples} represents the total number of medical reports selected.

In our experiments, the \textbf{Samples} value varied systematically from 500 to 10,000. From the experimental results, we observed a general positive trend in performance as the \textbf{Samples} value increased from 500 to 6,000, despite a slight downward trend between 1,000 and 3,000. The best performance was achieved when the \textbf{Samples} value reached 6,000. However, as the \textbf{Samples} value increased from 6,000 to 10,000, the results showed a continuous decline in performance. This indicates that the optimal model performance is achieved when \textbf{Samples} is set to 6,000.

\begin{table}
\centering
\caption{Compare the influence of different values of the important parameter \textbf{Beta} on the results in Hopfield Network.  }  
\label{tab:Hopfield_Network_Beta}
\small 
\begin{tabular}{c|cccc} 
\hline 
\toprule [0.5 pt] 
\textbf{Beta}  &\textbf{BLEU-4}  &  \textbf{ROUGE-L} & \textbf{METEOR} & \textbf{CIDEr} \\
\hline 
 0.5  & 0.126 & 0.281 & 0.161 & 0.217 \\
\hline 
 1.0 & 0.132 & 0.285 & 0.166 & 0.249 \\
\hline 
 2.0 & 0.130 & 0.284 & 0.161 & 0.226 \\
\hline 
 4.0 & \textbf{0.136} & \textbf{0.291} & \textbf{0.174} & 0.261\\
\hline 
 8.0 & 0.129 & 0.288 & 0.170 & \textbf{0.268} \\
\hline 
 16.0 & 0.129 & 0.283 & 0.162 & 0.221 \\
\hline 
\toprule [0.5 pt]  
\end{tabular}
\end{table}

\noindent $\bullet$ \textbf{Analysis of the key parameter \textit{Beta} in Hopfield.~}
The inverse temperature \textbf{Beta} is a crucial parameter in modern Hopfield networks. High values of \textbf{Beta} correspond to a low temperature, meaning the attraction basins of individual patterns remain separated, making the appearance of metastable states unlikely. Conversely, low values of \textbf{Beta} correspond to a high temperature, increasing the likelihood of forming metastable states. As shown in Table \ref{tab:Hopfield_Network_Beta}, we conducted a series of experiments with \textbf{Beta} values ranging from 0.5 to 16.0 to determine the optimal value. The results clearly indicate that the accuracy generally increases with \textbf{Beta} values from 0.5 to 4.0 (with a slight dip between 1.0 and 2.0). Beyond 4.0, the accuracy declines as expected. Therefore, we can preliminarily conclude that a \textbf{Beta} value of 4.0 yields the best performance for modern Hopfield networks in our experimental setup.

\begin{table}
\centering
\caption{Compare the impact of different report encoders on the results.  }  
\label{tab:Different_Encoders}
\small 
\resizebox{\linewidth}{!}{ 
\begin{tabular}{l|cccc} 
\hline 
\toprule [0.5 pt] 
\textbf{Encoder}  &\textbf{BLEU-4}  &  \textbf{ROUGE-L} & \textbf{METEOR} & \textbf{CIDEr} \\
\hline 
 Llama-2~\cite{touvron2023llama2}  & 0.129 & 0.285 & 0.166 & 0.234 \\
\hline 
 Vicuna-v1.5~\cite{zheng2023vicuna}  & 0.132 & 0.288 & 0.164 & 0.230 \\
\hline 
 Orca-2~\cite{mitra2023orca}  & 0.129 & 0.288 & 0.171 & 0.254 \\
\hline 
 InternLM~\cite{cai2024internlm2} & 0.129 & 0.290 & 0.171 & \textbf{0.267} \\
\hline 
 Bio\_ClinicalBERT~\cite{Emily2019Bio_ClinicalBERT}  & \textbf{0.136} & \textbf{0.291} & \textbf{0.174} & 0.261\\
\hline
\toprule [0.5 pt]  
\end{tabular}
}
\end{table}

\noindent $\bullet$ \textbf{Analysis of different text encoders.~}
Table~\ref{tab:Different_Encoders} shows the impact of using different text encoders on the results.  We explored the use of Llama2~\cite{touvron2023llama2}, Vicuna-v1.5~\cite{zheng2023vicuna}, Orca-2~\cite{mitra2023orca}, InternLM~\cite{cai2024internlm2}, and Bio\_ClinicalBERT~\cite{Emily2019Bio_ClinicalBERT} as report encoders to transform the reports into feature vectors.  From the table, it is evident that Bio\_ClinicalBERT~\cite{Emily2019Bio_ClinicalBERT} outperforms the other encoders overall.  We speculate that this is due to Bio\_ClinicalBERT~\cite{Emily2019Bio_ClinicalBERT} being fine-tuned specifically for medical data, giving it a distinct advantage in handling medical reports.

\begin{table}
\centering
\caption{Compare the impact of different CAM on the results.  }  
\label{tab:Different_CAMs}
\small 
\resizebox{\linewidth}{!}{ 
\begin{tabular}{l|cccc|ccc} 
\hline 
\toprule [0.5 pt] 
\multirow{2}{*}{\textbf{CAM Type}}  & \multicolumn{4}{c|}{\textbf{NLG Metrics}}   & \multicolumn{3}{c}{\textbf{CE Metrics}}  \\ \cline{2-8}
 &\textbf{B4}  &  \textbf{R-L} & \textbf{M} & \textbf{C}  & \textbf{P} & \textbf{R} & \textbf{F1} \\
\hline 
 HiResCAM & 0.131 & 0.284 & 0.164 & 0.242 & 0.530 & 0.414 & 0.465 \\
\hline 
 XGradCAM & 0.129 & 0.290 & 0.171 & 0.256 & 0.551 & 0.439 & 0.489 \\
\hline 
 EigenCAM & 0.126 & 0.281 & 0.160 & 0.203 & 0.512 & 0.324 & 0.397 \\
\hline 
 GradCAM++ & 0.127 & 0.289 & 0.170 & 0.251 & 0.553 & \textbf{0.459} & \textbf{0.502} \\
\hline 
 GradCAM & \textbf{0.136} & \textbf{0.291} & \textbf{0.174} & \textbf{0.261} & \textbf{0.555} & 0.429 & 0.484\\
\hline
\toprule [0.5 pt]  
\end{tabular}
} 
\end{table}

\noindent $\bullet$ \textbf{Analysis of different CAMs~\footnote{\url{https://github.com/jacobgil/pytorch-grad-cam}}.~}
Table \ref{tab:Different_CAMs} illustrates the impact of different activation map generation methods on the results. We explored five methods: 
HiResCAM, XGradCAM, EigenCAM, GradCAM++, and GradCAM. 
GradCAM is easy to implement and highly interpretable as it directly uses gradient information. GradCAM++ provides more accurate localization of different categories. HiResCAM produces high-resolution activation maps that capture finer details. XGradCAM comprehensively utilizes information from the network. EigenCAM generates more representative activation maps, aiding in a better understanding of the model's decision-making process. As shown in the table, GradCAM performs well on natural language generation (NLG) metrics, while GradCAM++ and XGradCAM, due to their ability to more accurately locate different categories and comprehensively utilize network information, respectively, show better overall performance on CE metrics compared to GradCAM. Overall, GradCAM excels in NLG metrics, while in terms of CE metrics, the differences among the five methods are minimal. After weighing the options, we believe GradCAM is more suitable.

\begin{table}
\centering
\caption{Compare the impact of different LLMs on the MIMIC-CXR dataset.   }
\label{tab:Different_LLMs}
\small 
\resizebox{\linewidth}{!}{ 
\begin{tabular}{l|cccc} 
\hline 
\toprule [0.5 pt] 
\textbf{LLMs}  &\textbf{BLEU-4}  &  \textbf{ROUGE-L} & \textbf{METEOR} & \textbf{CIDEr} \\
\hline 
 GPT2-Medium~\cite{radford2019gpt2} & 0.073 & 0.230 & 0.076 & 0.089 \\
\hline 
 InternLM-7B~\cite{cai2024internlm2} & 0.086 & 0.226 & 0.140 & 0.141 \\
\hline 
 Yi-1.5-9B~\cite{ai2024yi} & 0.106 & 0.280 & 0.136 & 0.149 \\
\hline 
 Orca-2-7b~\cite{mitra2023orca} & 0.129 & 0.289 & 0.168 & \textbf{0.262} \\
\hline 
 Vicuna-7b-v1.5~\cite{zheng2023vicuna} & 0.130 & 0.289 & 0.168 & 0.250 \\
\hline 
 Llama2-7B~\cite{touvron2023llama2} & \textbf{0.136} & \textbf{0.291} & \textbf{0.174} & \underline{0.261}\\
\hline
\toprule [0.5 pt]  
\end{tabular}
} 
\end{table}

\noindent $\bullet$ \textbf{Analysis of different LLMs.~}
In this section, we also evaluate the impact of different LLMs as report decoders on the final results.  As shown in Table \ref{tab:Different_LLMs}, we assessed a total of six LLMs: GPT2-Medium~\cite{radford2019gpt2}, InternLM-7B~\cite{cai2024internlm2}, Yi-1.5-9B~\cite{ai2024yi}, Orca-2-7b~\cite{mitra2023orca}, Vicuna-7b-v1.5~\cite{zheng2023vicuna}, and Llama2-7B~\cite{touvron2023llama2}. From the table, it is evident that Llama2-7B generally outperforms the other LLMs, with Orca-2-7b and Vicuna-7b-v1.5 also showing good performance, particularly Orca-2-7b leading in the CIDEr metric.  The remaining LLMs did not perform exceptionally well.
Thus, we can preliminarily conclude that Llama2-7B serves as the best report decoder in this experimental setup.  It is noteworthy that while Orca-2-7b excels in the CIDEr metric, its performance on other metrics is slightly inferior to Llama2-7B.  Vicuna-7b-v1.5 also demonstrated competitive performance across multiple metrics.  In contrast, GPT2-Medium and other LLMs did not meet expectations.  This highlights the importance of selecting an appropriate LLM as the report decoder to enhance the overall performance of the model.



\subsection{Visualization}

\begin{figure*}
    \centering
    \includegraphics[width=\linewidth]{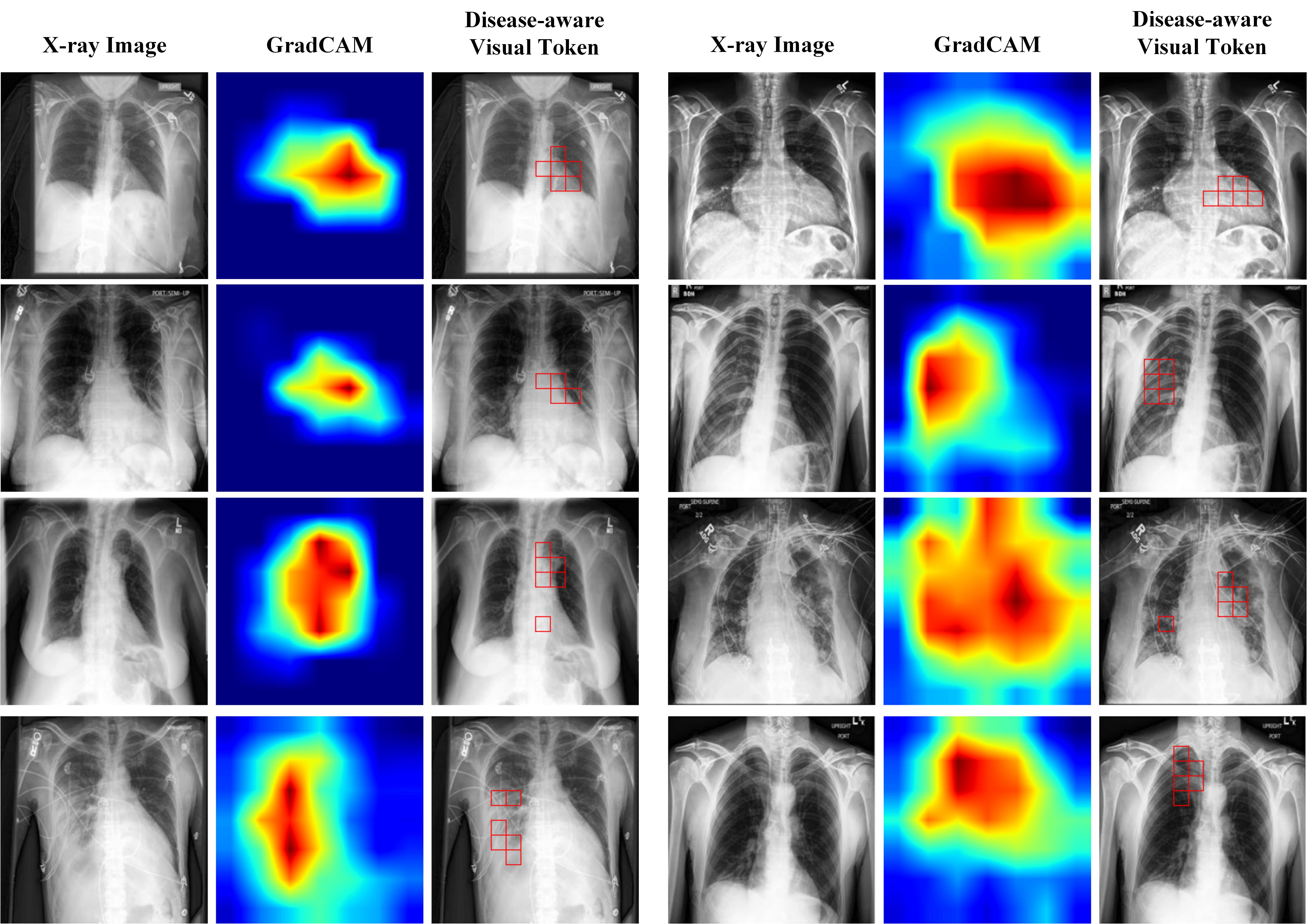}
    \caption{The gradcam activation maps and the mined visual tokens in stage 1.} 
    \label{fig:XrayCAM}
\end{figure*}

\begin{figure*}
    \centering
    \includegraphics[width=\linewidth]{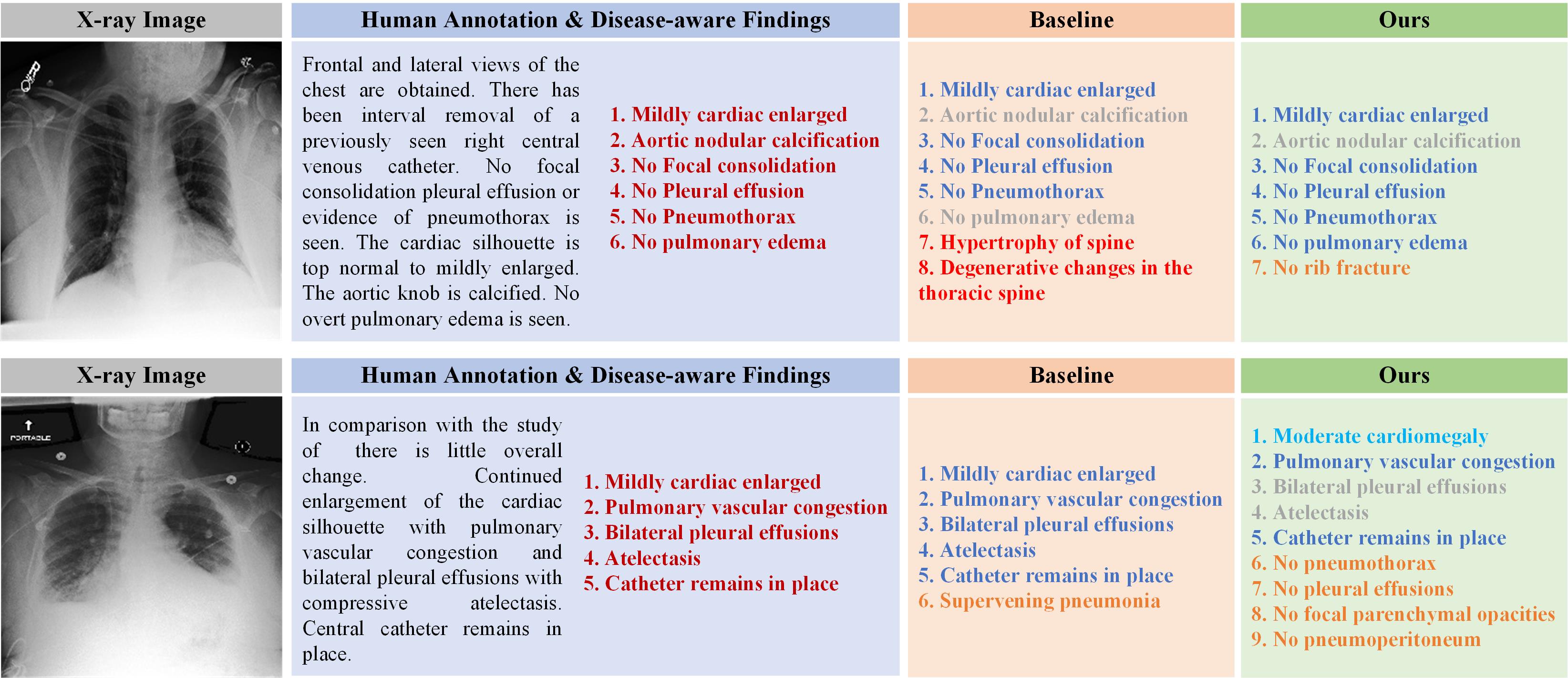}
    \caption{An illustration of the disease-aware findings of the baseline and our newly proposed model.} 
    \label{fig:diseaseFinding}
\end{figure*}

\newcommand{\MatchingAMGpt}[1]{\sethlcolor{pink}\hl{#1}}
\newcommand{\MatchingGpt}[1]{\sethlcolor{cyan}\hl{#1}}
\newcommand{\MatchingAM}[1]{\sethlcolor{yellow}\hl{#1}}

\begin{figure*}[h]
\centering
\resizebox{\textwidth}{!}{
\begin{tabular}{lp{7.4cm}p{7.4cm}p{7.4cm}}
\hline \toprule [0.5 pt] 
\multicolumn{1}{c}{\textbf{Image}} & \multicolumn{1}{c}{\textbf{Ground Truth}} & \multicolumn{1}{c}{\textbf{Ours}} & \multicolumn{1}{c}{\textbf{R2GenGPT}} \\ 
\hline \toprule [0.5 pt]

\begin{minipage}[b]{0.4\columnwidth}\vspace{7pt}\centering\raisebox{-.9\height}{\includegraphics[width=\linewidth]{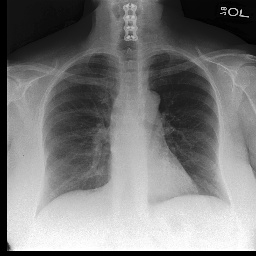}}\end{minipage} 
 & \MatchingAM{Pa and lateral views of the chest provided.} \MatchingAMGpt{Cervical spinal hardware} again noted. \MatchingGpt{Clips noted in the upper abdomen.} \MatchingAM{There is }\MatchingAMGpt{no}\MatchingAM{ focal consolidation effusion or} \MatchingAMGpt{pneumothorax. The cardiomediastinal silhouette is normal.} \MatchingAM{Imaged osseous structures are intact.} \MatchingAMGpt{No free air below the right hemidiaphragm}\MatchingAM{ is seen.}
 & \MatchingAM{Pa and lateral views of the chest provided. There is no focal consolidation effusion or pneumothorax. The cardiomediastinal silhouette is normal. Imaged osseous structures are intact. No free air below the right hemidiaphragm is seen. Cervical spinal hardware} is partially visualized in the lower cervical spine. 
 & The lungs are clear \MatchingGpt{without focal consolidation.} \MatchingGpt{No} pleural effusion or \MatchingGpt{pneumothorax} is seen. \MatchingGpt{The cardiac} and mediastinal \MatchingGpt{silhouettes are unremarkable}. \MatchingGpt{Cervical spinal} fusion \MatchingGpt{hardware} is partially imaged in the lower cervical spine. There is \MatchingGpt{no free air below the right hemidiaphragm.} Surgical \MatchingGpt{clips are noted in the upper abdomen.} \\ 

\begin{minipage}[b]{0.4\columnwidth}\vspace{7pt}\centering\raisebox{-.9\height}{\includegraphics[width=\linewidth]{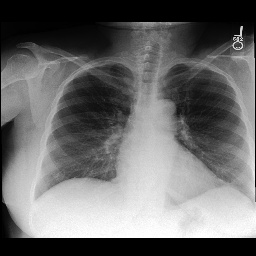}}\end{minipage} 
 & \MatchingAMGpt{Heart size is normal. The mediastinal and hilar contours are normal.} \MatchingAM{The pulmonary vasculature is normal.} \MatchingAMGpt{Lungs are clear. No pleural effusion or pneumothorax is seen.} \MatchingAM{There are no acute osseous abnormalities.}
 & \MatchingAM{Heart size is normal. The mediastinal and hilar contours are normal. The pulmonary vasculature is normal. Lungs are clear. No pleural effusion or pneumothorax is seen. There are no acute osseous abnormalities.} Surgical clips are noted in the right upper quadrant of the abdomen suggestive of prior cholecystectomy. 
 & The \MatchingGpt{heart is normal} in \MatchingGpt{size}. \MatchingGpt{The mediastinal and hilar contours} appear within \MatchingGpt{normal} limits. There is \MatchingGpt{no pleural effusion or pneumothorax}. The \MatchingGpt{lungs} appear \MatchingGpt{clear}. Bony structures are unremarkable. There is no free air below the right hemidiaphragm. Surgical clips are noted in the right upper quadrant. \\

\begin{minipage}[b]{0.4\columnwidth}\vspace{7pt}\centering\raisebox{-.9\height}{\includegraphics[width=\linewidth]{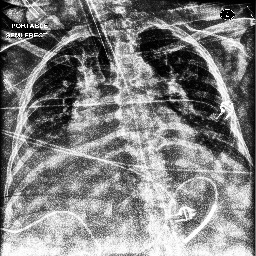}}\end{minipage} 
 & \MatchingAMGpt{As compared to the previous radiograph}\MatchingAM{ the patient has received a} new dobbhoff \MatchingAM{tube. The tip of the tube projects over the middle parts of the stomach. The course of the tube is unremarkable.} There is \MatchingAM{no evidence of complications notably}\MatchingAMGpt{ no pneumothorax.} Otherwise the radiographic appearance of the thoracic organs is similar to the previous examination.
 & \MatchingAM{As compared to the previous radiograph the patient has received a }nasogastric\MatchingAM{ tube. The tip of the tube projects over the middle parts of the stomach. The course of the tube is unremarkable. No evidence of complications notably no pneumothorax.} Moderate bilateral pleural effusions with underlying areas of atelectasis. Moderate cardiomegaly. No evidence of pulmonary edema. No focal parenchymal opacities suggestive of pneumonia.
 & \MatchingGpt{As compared to the previous radiograph} there is no relevant change. The monitoring and support devices are constant. Moderate cardiomegaly with bilateral pleural effusions of similar extent on the right and slightly larger on the left. Mild pulmonary vascular congestion. No new parenchymal opacities. \MatchingGpt{No pneumothorax.} Unchanged position of the endotracheal tube and the nasogastric tube which terminates in the stomach. \\

\begin{minipage}[b]{0.4\columnwidth}\vspace{7pt}\centering\raisebox{-.9\height}{\includegraphics[width=\linewidth]{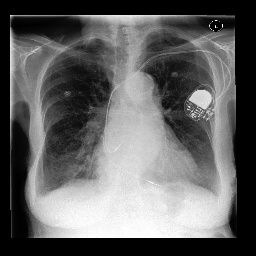}}\end{minipage}\vspace{7pt} 
 & \MatchingAM{Frontal and lateral views of the chest were obtained. Dual-lead }\MatchingAMGpt{left-sided pacemaker is again seen with leads}\MatchingAM{ extending to the expected positions of }\MatchingAMGpt{the right atrium and right ventricle.} The lungs are hyperinflated with flattening of the diaphragms suggesting chronic obstructive pulmonary disease. \MatchingAMGpt{No pleural effusion or pneumothorax is seen.} Slight increased opacity at the right lung base best seen on the fron.
 & \MatchingAM{Frontal and lateral views of the chest were obtained. Dual-lead left-sided pacemaker is again seen with leads extending to the expected positions of the right atrium and right ventricle.} There is \MatchingAM{no} focal consolidation \MatchingAM{pleural effusion or pneumothorax.} The cardiac silhouette is mildly enlarged. The aorta is calcified and tortuous. No acute osseous abnormalities are identified.
 & Pa and lateral views of the chest are compared to previous exam from. \MatchingGpt{Left-sided pacemaker is again seen with leads }terminating in \MatchingGpt{the right atrium and right ventricle.} There is \MatchingGpt{no} focal consolidation \MatchingGpt{pleural effusion or pneumothorax.} Cardiomediastinal silhouette is unchanged. Bony structures are intact. Atherosclerotic calcifications are noted at the aortic knob. \\ 

\hline \toprule [0.5 pt] 
\end{tabular}
}
\caption{X-ray images and their corresponding ground-truths, along with the output of our model and R2GenGPT model generation reports on the MIMIC-CXR dataset. Matching sentences in our report are highlighted in yellow, R2GenGPT matching sentences are highlighted in cyan, and sentences matching by both models are highlighted in pink.}
\label{fig:visual_report}
\end{figure*}

\noindent $\bullet$ \textbf{Disease-aware Visual Tokens.~} 
As illustrated in Fig.~\ref{fig:XrayCAM}, we have visualized the mechanism by which our AM-MRG model extracts disease-aware visual tokens. We have showcased a collection of eight chest X-ray images, each accompanied by the original scan, the activation maps rendered through GradCAM, and the highlighted regions superimposed on the X-rays. These visualizations demonstrate the AM-MRG model's proficiency in pinpointing the exact locations of significant diseases and isolating the pertinent regions for further model processing.
In instances where X-rays exhibit expansive lesion areas with numerous activated regions, we focus on the top six regions exhibiting the highest activation scores. Conversely, for X-rays with more confined lesion areas and a smaller number of activated regions, we apply a threshold to retain only those regions with activation values exceeding it. This strategy prioritizes the quality of the selected visual regions over their quantity, thereby ensuring their reliability and stability.
This refined approach heightens the model's sensitivity to crucial lesion areas, enhancing the precision of the resultant reports. It guarantees the production of comprehensive reports that are rich in interpretability and detail accuracy. Furthermore, by adeptly pinpointing and extracting key lesion areas, our method excels in handling a diverse range of chest X-rays, from the intricate to the straightforward, accurately capturing and representing the underlying disease state.

\noindent $\bullet$ \textbf{Disease Finding.~} 
As shown in Fig.~\ref{fig:diseaseFinding}, to demonstrate the effectiveness of our AM-MRG model in extracting disease-related information, we compared two chest X-rays and their corresponding reports to highlight the differences between the AM-MRG and Baseline model.  For each medical report pair, we displayed the original image, the corresponding ground truth, and the disease entities extracted from the ground truth.  We also extracted disease entities from the predicted reports generated by the Baseline and AM-MRG model. In this figure, dark red indicates the true disease entities extracted from the ground truth, blue represents the disease entities in the predicted report that match the ground truth, light blue indicates entities in the predicted report that match the ground truth but with slight differences in the description, gray shows the entities present in the ground truth but not predicted by the model (i.e., missed by the model),  bright red indicates entities predicted by the model but not present in the ground truth (i.e., false positives), and brownish-yellow represents normal conditions predicted by the model but not mentioned in the ground truth. 

From the comparison, it is evident that the AM-MRG model significantly outperforms the Baseline model in capturing and describing disease entities.  The AM-MRG model not only more accurately identifies key diseases but also reduces the number of false predictions and avoids missing critical disease information.  This indicates that the AM-MRG model can better reflect the actual condition of the patient when generating medical reports, providing clinicians with more reliable diagnostic information.

\noindent $\bullet$ \textbf{Generated Reports.~}
As shown in Fig.~\ref{fig:visual_report}, we present several examples to demonstrate the effectiveness of our proposed AM-MRG model for X-ray image-based report generation.   For specific X-ray images, we compared the ground truth reports with those generated by the AM-MRG model and the R2GenGPT~\cite{Wang2023R2GenGPT} model, allowing for a more comprehensive and rational evaluation. To provide a clearer visualization, we have highlighted the parts of the generated reports that match the ground truth.   The yellow highlights indicate sections of the report generated by our model that align with the ground truth, the blue highlights indicate sections of the report generated by the R2GenGPT~\cite{Wang2023R2GenGPT} model that match the ground truth, and the pink highlights represent sections that both models accurately captured.

It is evident that the report generated by our model is more closely aligned with the actual report compared to the one generated by the R2GenGPT~\cite{Wang2023R2GenGPT} model.   This highlights the effectiveness of our AM-MRG model in generating accurate and reliable X-ray image reports.

\subsection{Limitation Analysis} 
Owing to the deployment of large language models, the training of our model necessitates the use of high-end GPU graphics cards to facilitate the training process. The pursuit of further optimization and acceleration of the model, such as through quantization techniques, is an imperative challenge that demands immediate attention. Moreover, we have incorporated a simple report memory mechanism to bolster the ultimate performance of the model. However, this method does not undergo a thorough analysis of the reports, thereby not fully leveraging the potential insights contained within the annotated reports.

\section{Conclusion and Future Works}   \label{sec::conclusion}
In this paper, we propose a new framework for the generation of X-ray based medical reports, termed AM-MRG. The key idea of this work is to exploit the disease-aware vision tokens and report memories for high-performance report generation. We first obtain the vision tokens based on disease classification network and build a vision memory bank. Then, we introduce modern Hopfield networks to get vision memories and report memories for the large language model based X-ray medical report generation. Extensive experiments on three benchmark datasets fully validated the effectiveness of our proposed AM-MRG framework. 
In future work, we plan to integrate vision tokens into the medical knowledge graph and utilize this graph to guide the generation of medical reports.

\small{ 
\bibliographystyle{IEEEtran}
\bibliography{reference}
}

\end{document}